\documentclass{article}

\usepackage{PRIMEarxiv}

\usepackage[utf8]{inputenc} 
\usepackage[T1]{fontenc}    
\usepackage{hyperref}       
\usepackage{url}            
\usepackage{booktabs}       
\usepackage{amsfonts}       
\usepackage{nicefrac}       
\usepackage{microtype}      
\usepackage{lipsum}
\usepackage{graphicx}
\graphicspath{{media/}}     

\usepackage{caption}
\usepackage{subcaption}
\usepackage{amsmath} 

\usepackage{xcolor}
  
\title{Evidence Triangulation for Multimodal Fact-Checking in the Wild\thanks{Accepted at the 2026 European Conference on Computer Vision (ECCV).}}

\author{
\textbf{Stefanos-Iordanis Papadopoulos}$^{1,2}$\thanks{Corresponding author} \quad
\textbf{Zacharias Chrysidis}$^{1}$ \quad
\textbf{Christos Koutlis}$^{1}$ \\
\textbf{Symeon Papadopoulos}$^{1}$ \quad
\textbf{Panagiotis C. Petrantonakis}$^{2}$ \\
\\
$^{1}$Information Technologies Institute, Centre for Research \& Technology, Hellas, Greece\\
$^{2}$Department of Electrical and Computer Engineering, Aristotle University of Thessaloniki, Greece\\
\\
\{stefpapad,zchrysid,ckoutlis,papadop\}@iti.gr, ppetrant@ece.auth.gr
}

\begin{document}
\maketitle

\begin{abstract}
The proliferation of multimedia content on social platforms has fueled multimodal misinformation, where images are used to reinforce false claims. Consequently, Multimodal Fact-Checking (MFC) has emerged as an increasingly important research area. 
However, current progress is hindered by a reliance on synthetic training data and curated benchmarks that fail to capture the complexity of in-the-wild data. 
Furthermore, existing detection models rely on restricted intra-modality consistency or unconstrained all-to-all fusion, failing to capture nuanced relations between posts and external evidence. 
To address these limitations, we introduce X-POSE, a benchmark of real-world, community-annotated multimodal posts from X (formerly Twitter), augmented with full-length news articles retrieved via VLM-optimized search.
Additionally, we propose TRENT, a novel MFC model that performs evidence triangulation using three parallel cross-attention streams alongside a relational fusion mechanism that explicitly models entailment and contradiction. 
Extensive evaluations demonstrate that TRENT consistently outperforms state-of-the-art specialized models and commercial VLMs.
The code, prompt templates, and dataset are available at
\url{https://github.com/stevejpapad/evidence-triangulation}.
\end{abstract}


\section{Introduction}
\label{sec:intro}

Digital information ecosystems are increasingly defined by multimodal content. 
Videos, images, text, and synthetic media enable richer communication and expression, but also introduce new forms and avenues for large-scale deception. 
Beyond edited and AI-generated images, a key challenge is multimodal misinformation, in which images are paired with false claims to increase their perceived credibility \cite{newman2020non} and wide-spread diffusion \cite{li2020picture}.
In response, researchers have been developing automated Multimodal Fact-Checking (MFC) systems to retrieve relevant information from the Web to cross-examine and verify the veracity of multimodal claims \cite{akhtar2023multimodal}.

\begin{figure}[t]
    \centering
    \begin{subfigure}{\linewidth}
        \centering
        \includegraphics[width=\linewidth]{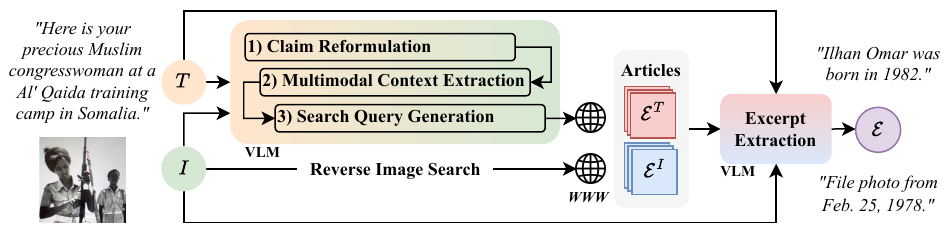}
        \caption{}
        \label{fig:pipeline}
    \end{subfigure}

    \begin{subfigure}{\linewidth}
        \centering
        \includegraphics[width=\linewidth]{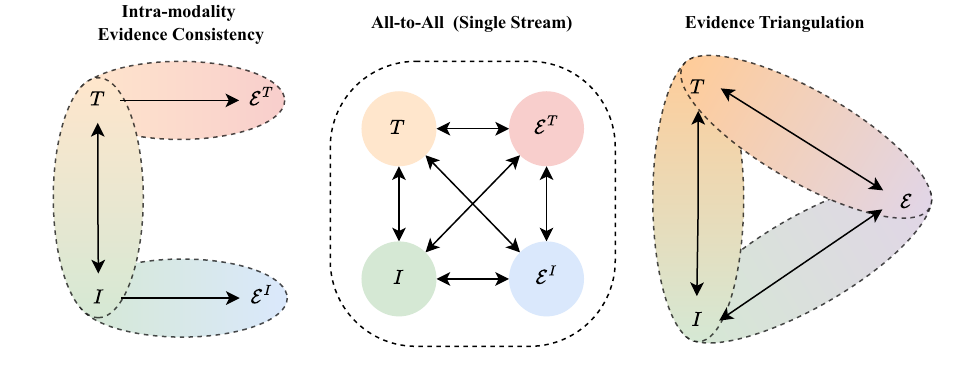}
        \caption{}
        \label{fig:comparison}
    \end{subfigure}
    \caption{
    (a) Proposed evidence collection pipeline: A VLM reformulates a multimodal post $(I, T)$ into search queries to retrieve relevant articles ($\mathcal{E}^T, \mathcal{E}^I$) and extract relevant excerpts ($\mathcal{E}$).
    (b) Conceptual comparison of MFC architectures: Evidence Triangulation leverages dedicated streams to explicitly isolate the three-way relationship between image, text, and evidence, instead of relying on intra-modality or all-to-all interactions.
    }
    \label{fig:stacked}
\end{figure}

To date, developing robust MFC models for image-text verification is constrained by the scarcity of high-quality data. As a result, recent work relies on weakly annotated or synthetic datasets~\cite{mishra2022factify, luo2021newsclippings, aneja2023cosmos, biamby2022twitter, sabir2018deep, muller2020multimodal}, or benchmarks sourced directly from fact-checking organizations~\cite{papadopoulos2024verite, cao2026averimatec, rothermel2026veritas}. 
The former introduces label noise, and synthetic data fail to capture the visual, linguistic, and cultural nuances of \emph{in-the-wild} misinformation. 
Conversely, although datasets derived from fact-checking sources offer high-quality annotations, they provide clean, curated, and self-contained claims that do not reflect the noisy, contextual way users communicate on social media platforms.

To address this, we introduce X-POSE, a new in-the-wild, evidence-enhanced dataset 
of multimodal user posts from X (formerly Twitter), annotated through the Community Notes system\footnote{\url{https://communitynotes.x.com/guide}}, enabling both training and evaluation of MFC models---unlike prior work that used it only for evaluation \cite{xiao2025xfacta}.
We leverage VLMs to refine search queries, use both text-based and reverse image search APIs to retrieve full-length news articles as potential evidence, and apply VLMs to extract the article excerpts most relevant for supporting or refuting each claim. 
Moreover, we integrate source credibility to enhance evidence quality and community ratings to assess consensus for more reliable evaluation.

As shown
in Figure~\ref{fig:pipeline}, a post falsely claims a ``Muslim congresswoman'' was photographed training at an ``Al Qaeda camp,'' exploiting harmful religious stereotypes to incite distrust and hostility.
The VLM extracts a structured claim and contextual metadata (e.g., named entities, OCR text, image descriptions, and the ``5Ws'': Who, What, Where, When, Why), then generates search queries that, together with reverse image search, return relevant web articles. 
Finally, the VLM identifies the most pertinent evidence excerpts---
revealing that the inferred individual was born in 1982, while the photo was taken in 1978---providing the key information needed for the MFC system to dispute the claim.

On the modeling side, existing methods fall into three categories: 
(1) \textit{Intra-modality evidence consistency} models, which use separate streams to examine text-text evidence, image-image evidence, and the internal consistency of the image-text pair~\cite{abdelnabi2022open, zhang2023ecenet}; 
(2) \textit{All-to-All} models, which jointly attend over all modalities and evidence~\cite{papadopoulos2025red, Papadopoulos_2025_WACV}; and 
(3) \textit{VLM-based} methods~\cite{wu2025e2lvlm, braun2025defame}.
However, the first ignores cross-modal evidence relations, the second can struggle to identify subtle relations, while the third can suffer from high computational complexity.

To address these limitations, we introduce TRENT, an architecture that, as illustrated in Figure~\ref{fig:comparison}, models text-to-all evidence, image-to-all evidence, and image-to-text alignment using three parallel cross-attention streams. 
The resulting outputs are processed by a relational fusion mechanism designed to capture entailment and contradiction.
A comparative evaluation against commercial VLMs (including Grok 4, Claude Sonnet 4.6, Gemini 3, and GPT-5.4) and state-of-the-art (SotA) specialized MFC methods trained on X-POSE shows that TRENT consistently outperforms competing methods. 
Extensive ablation studies demonstrate that each component---evidence source credibility, evidence reranking, VLM-assisted excerpt extraction, and especially triangulated cross-attention and relational fusion---contributes significantly to overall performance.

\section{Background and Related Work}
\label{sec:rw}

\subsection{Problem Formulation}
\label{sec:problem}

Let $\mathcal{D} = \{(I_n, T_n, \mathcal{E}^{I}_n, \mathcal{E}^{T}_n, y_n)\}_{n=1}^{N}$ be a dataset of $N$ multimodal instances. 
Each instance consists of a multimodal pair $(I_n, T_n)$, where $I_n \in \mathcal{I}$ is an image and $T_n \in \mathcal{T}$ is its associated text (user post).
To verify the veracity of the pair, we incorporate two sets of external evidence:
\begin{itemize}
\item $\mathcal{E}^{I}_n = \{E^{I}_{n,1}, E^{I}_{n,2}, \ldots, E^{I}_{n,M}\}$ represents external information 
retrieved via reverse-image search using $I_n$.
\item $\mathcal{E}^{T}_n = \{E^{T}_{n,1}, E^{T}_{n,2}, \ldots, E^{T}_{n,M}\}$ represents external information  retrieved using the text $T_n$ as a query.
\end{itemize}
Each evidence set contains up to $M$ items (e.g., full-length news articles or article excerpts). 
The ground-truth label is denoted by $y_n \in \{0, 1\}$, where $y_n = 0$ represents a truthful claim and $y_n = 1$ indicates misinformation. 
The task of MFC is to learn a 
predictive function $f: (\mathcal{I}, \mathcal{T}, \mathcal{E}^I, \mathcal{E}^T) \rightarrow \hat{y}$, which minimizes the discrepancy between the predicted verdict $\hat{y}$ and the ground-truth $y$.

\subsection{Multimodal Fact-Checking Datasets}
\label{sec:mfc_datasets}

Due to the limited scale and event diversity of early MFC datasets \cite{zlatkova2019fact, jin2016novel, boididou2015verifying, boididou2018verifying}, subsequent works turned to scalable synthetic data generation via \emph{decontextualization} \cite{aneja2023cosmos, biamby2022twitter, luo2021newsclippings}, \emph{named-entity manipulation} \cite{sabir2018deep, muller2020multimodal, papadopoulos2023synthetic}, VLMs \cite{papadopoulos2025latent}, or automated weak supervision \cite{nakamura2020fakeddit}. 
To bridge the resulting distribution gap between algorithmically generated data and real-world misinformation, recent works leverage real-world evaluation benchmarks such as \textit{COSMOS} \cite{aneja2023cosmos} and \textit{VERITE} \cite{papadopoulos2024verite}. 
However, these benchmarks comprise self-contained claims curated by fact-checkers rather than uncurated user posts, restricting their suitability for studying misinformation in the wild.

Beyond internal image-text consistency, robust multimodal fact-checking typically necessitates external information to cross-examine claims. However, existing evidence-enhanced datasets suffer from significant limitations in scale, balance, and realism. 
The NewsCLIPpings+ dataset \cite{abdelnabi2022open} relies on synthetic decontextualization of news media and restricts textual evidence to article titles and brief captions. 
This setup frequently surfaces original or near-identical articles during evidence retrieval, causing data leakage \cite{glockner2022missing, Papadopoulos_2025_WACV}. 
Similarly, FACTIFY \cite{mishra2022factify} restricts its scope to official news handles on Twitter and relies on heuristic annotations, thus allowing models to exploit shortcuts and reach an inflated 99--100\% accuracy on the \textit{Refute} class \cite{suryavardan2023findings, chrysidis2024credible}, limiting its utility and realism.

Conversely, although \textit{AVerImaTeC} \cite{cao2026averimatec} and \textit{M4FC} \cite{geng2025m4fc} provide curated multimodal claims from fact-checked articles, their limited scale and class imbalances (e.g., only 17 `Supported' claims in \textit{AVerImaTeC}, 292 in \textit{M4FC}) prevent effective training.
Similarly, while \textit{VeriTaS} \cite{rothermel2026veritas} expands MFC beyond image-text pairs, it is intended for evaluation and comprises curated self-contained claims from fact-checking sites rather than user-generated content.

\begin{table}[t]
\centering
\caption{Comparison of multimodal fact-checking datasets.}
\label{tab:dataset_comparison}
\begin{tabular}{lccccc}
\toprule
\textbf{Dataset} & \textbf{Samples} & \textbf{Train} & \textbf{Eval.} & \textbf{Claims} & \textbf{Annotation} \\

\midrule

NewsCLIPpings+ \cite{abdelnabi2022open} & 85K & \checkmark & \checkmark (Synth.) & Synthetic & Automated \\

MMFakeBench \cite{liu2025mmfakebench} & 11K & -- & \checkmark (Synth.) & Synthetic & Automated \\

FACTIFY \cite{mishra2022factify} & 50K & \checkmark & \checkmark & News Handles & Automated\\

VERITE \cite{papadopoulos2024verite} & 1K & -- & \checkmark & Self-contained & Fact-checks \\

VeriTaS \cite{rothermel2026veritas} & 25K & -- & \checkmark & Self-contained & Fact-checks \\

M4FC \cite{geng2025m4fc} & 7K & \checkmark (Limited) & \checkmark & Self-contained & Fact-checks \\

AVerImaTeC \cite{cao2026averimatec} & 1.3K & \checkmark (Limited) & \checkmark & Self-contained & Fact-checks \\

X-FACTA \cite{xiao2025xfacta} & 2.4K & -- & \checkmark & In-the-wild & Crowdsourced \\

\midrule

\textbf{X-POSE (Ours)} & 5.7K & \checkmark & \checkmark & In-the-wild & Crowdsourced \\

\bottomrule
\end{tabular}
\end{table}

\subsection{Crowdsourced Fact-Checking}

Recent work has examined the effectiveness of crowdsourced fact-checking systems such as X’s Community Notes, which append clarifying annotations to potentially misleading posts. Adding a note has been shown to significantly reduce user engagement with misleading content \cite{slaughter2025community}; furthermore, posts labeled as misleading are much more likely to be voluntarily deleted by their authors \cite{gao2025can}. Notes citing neutral, high-quality sources also receive higher helpfulness scores, suggesting that contributors actively prioritize factual accuracy \cite{kangur2026checks}. Beyond human curation, recent research has explored generating notes with LLMs and predicting their subsequent helpfulness ratings \cite{singhlimitations, franzmeyer2024hellofresh, xing2025communitynotes, wu2025beyond}. Community Notes have also served as specialized benchmarks for the intent-aware classification of AI-generated images \cite{skoularikis2025humor} and multimodal fact-checking \cite{xiao2025xfacta}. 
However, they have not yet been utilized for both training and evaluating MFC models.

In contrast, as shown in Table~\ref{tab:dataset_comparison}, the proposed X-POSE dataset comprises uncurated, user-generated content of sufficient scale and balance to enable both the training and evaluation of MFC models under realistic, in-the-wild settings.

\subsection{Evidence-Based Detection}

Existing MFC methods can be classified into three paradigms. 
The first assesses intra-modality consistency (e.g., text-to-text and image-to-image evidence examination) alongside internal image-text consistency, leveraging memory networks, stance extraction modules, or named-entity co-occurrence features \cite{abdelnabi2022open, yuan2023support, zhang2023ecenet}. 
The second paradigm models all-to-all relationships across the claim,  image, and retrieved evidence blocks using dedicated attention- and Transformer-based architectures \cite{papadopoulos2023synthetic, papadopoulos2025red, Papadopoulos_2025_WACV}. 
However, approaches that restrict comparisons to intra-modality consistency often overlook vital cross-modal evidence relations, whereas full all-to-all architectures can capture broad salient cues but fail to isolate the subtle relations of entailment and contradiction.
More recently, a third paradigm explores VLMs and LLM-based agentic frameworks to rerank retrieved evidence and generate final verifications \cite{tahmasebi2024multimodal, qi2024sniffer, wu2025e2lvlm, lakara2024mad, braun2025defame, yan2025trust}. 

In this study, we synthesize these paradigms by leveraging VLMs for claim reformulation, evidence collection, and filtering, coupled with TRENT, a specialized, lightweight MFC architecture that explicitly models the relationships between internal image-text pairs, image-to-all evidence, and text-to-all evidence.

\section{Construction of X-POSE}
\label{sec:dataset}

\textbf{\textit{Data Collection and Annotation.}}
To construct X-POSE (\textit{X Posts with Evidence}), we source from the X Community Notes archive. From the \textit{Notes} file, we define the \emph{Truthful} class using notes marked as \textit{factually correct}, \textit{not satire}, \textit{not outdated}, and citing \textit{trustworthy sources}. For the \emph{Misinformation} class, we retain notes marked as \textit{missing important context} or containing \textit{factual errors} that still cite \textit{trustworthy sources}. We exclude cases involving manipulated or AI-generated images to focus on misinformation involving authentic but miscaptioned or out-of-context images. 
While identifying synthetic media is increasingly relevant to multimodal misinformation \cite{chrysidis2026synthetic}, it often demands distinct visual forensic analysis~\cite{mehrjardi2023survey, deng2025survey} that falls outside the scope of this study.

We collect the associated posts, retaining only multimodal samples that contain both text and images.
The final X-POSE comprises 5,704 unique image-text pairs, balanced across factually correct (2,881) and misinformation (2,823) samples, spanning 2017--2025. 
Averaging the outputs of two language detection tools---Fast-LangDetect and Lingua---we estimate that the majority of posts are in English (84.8\%), followed by French (3.5\%), Portuguese (2.1\%), Spanish (1.8\%), and German (1.2\%). Using the IPTC taxonomy\footnote{\url{https://www.iptc.org/std/NewsCodes/treeview}} with Gemma~3, we find X-POSE is primarily composed of political content (24.8\%), followed by arts and media (19.7\%), conflict and war (15.8\%), crime and justice (7.6\%), science and technology (6.8\%), and society (5.7\%), with the remaining 11 categories (e.g., health, economy, sports) each accounting for less than 5\%, reflecting the dataset's topical diversity.

\paragraph{\textbf{Consensus-based Data Filtering.}}
Crowdsourced annotations can be noisy due to annotator disagreements regarding note phrasing, source credibility, or subjective interpretations. 
To mitigate this, we filter examples using a \emph{helpfulness} score, which serves as a proxy for community consensus. 
For a given sample $n$, let $\eta_n^+$ and $\eta_n^-$ denote the number of \textsc{Helpful} and \textsc{Not Helpful} votes, respectively, aggregated from the raw Community Notes \texttt{ratings} files. An item's helpfulness score $h_n \in [0, 100]$ is defined as:
\begin{equation}
h_n =
\begin{cases}
\frac{\eta_n^+}{\eta_n^+ + \eta_n^-} \times 100, & \text{if } \eta_n^+ + \eta_n^- > 0,\\
0, & \text{otherwise.}
\end{cases}
\end{equation}
We retain samples where $h_n \ge \theta$ to construct high-agreement evaluation subsets. 
For simplicity, we refer to these as the $h \ge \theta$ subsets.

\paragraph{\textbf{Evidence Collection.}}
Our evidence collection pipeline is grounded in the workflows of professional fact-checkers, who (i) disambiguate claims, (ii) contextualize them by identifying relevant entities and circumstances, and (iii) formulate targeted queries to search for corroborating or contradicting evidence~\cite{sundriyal2023chaos, warren2025show, micallef2022true}. 
We model these steps using a three-stage VLM-assisted pipeline\footnote{See \texttt{src/vlm\_prompts.py} in the codebase for all prompt templates.}
(1)~\textbf{Claim Reformulation (\textit{Prompt 1})} transforms raw, noisy user posts into clear, self-contained textual claims. 
(2)~\textbf{Multimodal Context Extraction (\textit{Prompt 2})} extracts structured metadata, including named entities, image descriptions, OCR text, and the ``5Ws''. 
(3)~\textbf{Search Query Generation (\textit{Prompt 3})} produces concise queries for search APIs, conditioned on the reformulated claim, image, and extracted context.
We utilize Gemma~3 due to its consistent performance and lack of the restrictive safety filters---such as those regarding public figures or sensitive political discourse---frequently encountered in large proprietary models. 
We submit the generated queries and images to the Google Search and Lens APIs via ScrapingDog, retrieving up to 10 results per query.
We fetch the pages, extract clean body text and discard cookie-consent popups and JavaScript-rendering errors.

\paragraph{\textbf{Evidence Refinement.}}
To reduce unreliable content and prevent information leakage (e.g., retrieving the associated Community Note), we exclude major social media platforms (X, Facebook, Instagram, YouTube, and TikTok) and use the MBFC\footnote{\url{https://github.com/drmikecrowe/mbfcext}} taxonomy to assess source credibility as a proxy for evidence quality and remove low-credibility outlets.
From 44,559 retrieved pages, we remove 6,366 social media items and 1,404 low-credibility articles, while 18,536 remain unrated. 
The final collection comprises articles from 10,296 unique domains, with the most frequent sources including Wikipedia (1,958), BBC (704), The Guardian (680), CNN (507), Al Jazeera (403), and NPR (341).
Furthermore, because 49.9\% of these collected articles postdate the original user post, we conduct additional experiments under a past-only evidence setting to more accurately simulate the fact-checking of novel claims without future data leakage. 

\paragraph{\textbf{Excerpt Extraction.}}
Information relevant to a specific claim is typically concentrated in localized segments of full-length articles \cite{vladika2023scientific}. Accordingly, we utilize a VLM, MiniCPM-V 2.6\footnote{\url{https://huggingface.co/openbmb/MiniCPM-V-2_6}} \cite{yao2024minicpm}, to perform cross-modal reasoning between the multimodal claim (text and image) and each retrieved article. Using \textit{Prompt 4}, the VLM extracts \textit{verbatim excerpts} from the text that are directly relevant to the claim. This process yielded 22,226 evidence excerpts (median length: 61 words). Of the collected posts, 4,785 are associated with at least one relevant excerpt, while 919 yielded no relevant evidence. The number of excerpts per post ranges up to a maximum of 16, with a median of 4; 79.4\% originate from text-based retrieval and 20.6\% from image-based retrieval.

\section{The TRENT Architecture}
\label{sec:method}

\subsection{Multimodal Representation}

We employ CLIP ViT-L/14~\cite{radford2021learning} as a pre-trained encoder to extract image embeddings $\mathbf{i} \in \mathbb{R}^d$ and text embeddings $\mathbf{t} \in \mathbb{R}^d$, where $d = 768$ denotes the embedding dimension. 
Similarly, the top-$M$ retrieved evidence excerpts are encoded into matrices $\mathbf{e}^{I}, \mathbf{e}^{T} \in \mathbb{R}^{d \times M}$ representing image- and text-based evidence, respectively, or concatenated into a single matrix $\mathbf{e} \in \mathbb{R}^{d \times 2M}$ when combined.

\subsection{Evidence Reranking}

Search engine APIs already include mechanisms to retrieve and rank potentially relevant content based on a given image-text pair. Nevertheless, as prior studies have shown \cite{papadopoulos2025red}, it is often beneficial to apply a 
reranking step. 
For each post $n$, we rerank retrieved evidence items using cosine similarity in the embedding space. Specifically, we compare the claim text embedding $\mathbf{t}_n$ against all text-retrieved evidence embeddings $\mathbf{e}_n^T$, and the claim visual embedding $\mathbf{i}_n$ against all image-retrieved evidence embeddings $\mathbf{e}_n^I$. 
Each evidence set contains up to 10 items. 
We then select the top-$M$ evidence items with the highest similarity scores. 
If fewer than $M$ candidates are available, we pad with zero vectors. 

\subsection{Evidence Triangulation}

\begin{figure}[t]
    \centering
    \includegraphics[width=\linewidth]{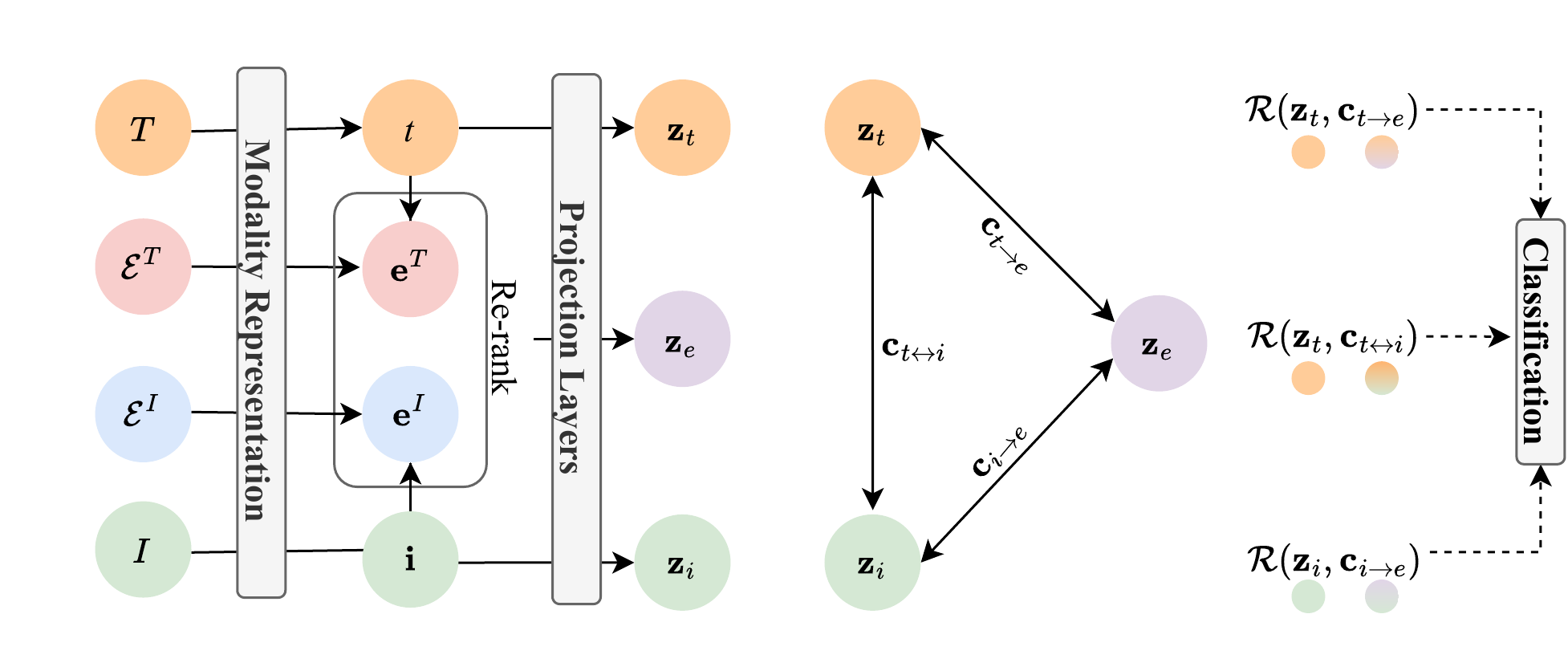} 
    \caption{Overview of TRENT: Input text ($T$), image ($I$), and evidence excerpts ($\mathcal{E}$) are projected to embeddings ($\mathbf{z}_t, \mathbf{z}_i, \mathbf{z}_e$) and processed by three parallel cross-attention streams. 
    The resulting contextualized representations ($\mathbf{c}_{t \rightarrow e}, \mathbf{c}_{t \rightarrow i}, \mathbf{c}_{i \leftrightarrow e}$) are integrated via relational fusion $\mathcal{R}(\cdot)$ to capture entailment and contradiction for classification.}    
    \label{fig:TRENT}
\end{figure}

TRENT employs three parallel cross-attention transformer blocks $\mathrm{C(\cdot)}$ and relational fusion $\mathcal{R(\cdot)}$ to capture interactions across images, texts, and evidence.

First, raw embeddings are projected into a shared latent space of dimension $d_c = 512$ using modality-specific learnable linear projections:
\begin{equation}
    \mathbf{z}_i = \mathbf{W}_i \ \mathbf{i}, \quad \mathbf{z}_t = \mathbf{W}_t \ \mathbf{t}, \quad \mathbf{z}_e = \mathbf{W}_e \ \mathbf{e}
\end{equation}
where $\mathbf{W}_i, \mathbf{W}_t, \mathbf{W}_e \in \mathbb{R}^{d_c\times d}$ for image, text, and evidence, respectively. 

Given query $\mathbf{z}_q$ and key-value $\mathbf{z}_{kv}$ vectors, a $\mathrm{C(\cdot)}$ block is defined as:
\begin{align}
    \mathbf{a} &= \text{LN}(\mathbf{z}_q + \mathrm{MHA}(\mathbf{z}_q, \mathbf{z}_{kv}, \mathbf{z}_{kv})) \\   
    \mathrm{C}(\mathbf{z}_q, \mathbf{z}_{kv}) &= \text{LN}(\mathbf{a} + \text{FFN}(\mathbf{a}))    
\end{align}
where $\mathrm{MHA}(\cdot)$ denotes multi-head attention with $h=8$ heads and a key padding mask to ignore zero-padded entries,  
$\text{FFN}(\mathbf{z}) = \mathbf{W}_2(\text{GELU}(\mathbf{W}_1 \mathbf{z}))$ is a Feed-Forward Network with $\mathbf{W}_1, \mathbf{W}_2\in\mathbb{R}^{d_c\times d_c}$, 
and Layer Normalization (LN) is applied after each residual connection.

We implement three parallel verification streams to examine the relationship between: (i) the visual content and all external evidence ($\mathbf{c}_{i \rightarrow e}$), (ii) the textual claim and all external evidence ($\mathbf{c}_{t \rightarrow e}$), and (iii) the internal cross-modal consistency between the paired text and image ($\mathbf{c}_{t \leftrightarrow i}$). 
Formally expressed:
\begin{align}
\label{eq:c}
    \mathbf{c}_{i \rightarrow e} = \mathrm{C}(\mathbf{z}_i, \mathbf{z}_e), \
    \mathbf{c}_{t \rightarrow e} = \mathrm{C}(\mathbf{z}_t, \mathbf{z}_e), \
    \mathbf{c}_{t \leftrightarrow i} &= \mathrm{C}(\mathbf{z}_t, \mathbf{z}_i)
\end{align}

\paragraph{\textbf{Relational Fusion.}}
To capture entailment and contradiction relations between the image-text pair and the external evidence, we employ a relational fusion function $\mathcal{R(\cdot)}$ inspired by standard formulations in natural language inference (NLI) \cite{conneau2017supervised}, but adapted for MFC:
\begin{equation}
    \mathcal{R}(\mathbf{z}_1, \mathbf{z}_2) = [ \mathbf{z}_1 ; \mathbf{z}_2 ; |\mathbf{z}_1 - \mathbf{z}_2| ; \mathbf{z}_1 \odot \mathbf{z}_2 ]
\end{equation}
where $[;]$ denotes concatenation and $\odot$ denotes the element-wise product. 
Applying $\mathcal{R}$ to each stream and concatenating yields the final representation: 
\begin{equation}
\label{eq:z}
    \mathbf{z}_{r} = [\mathcal{R}(\mathbf{z}_i, \mathbf{c}_{i \rightarrow e}) ; 
    \mathcal{R}(\mathbf{z}_t, \mathbf{c}_{t \rightarrow e}) ; 
    \mathcal{R}(\mathbf{z}_t, \mathbf{c}_{t \leftrightarrow i})]
\end{equation} 
with $\mathbf{z_r} \in\mathbb{R}^{12d_c}$; $4d_c$ per stream.

\paragraph{\textbf{Classification.}}
The final veracity score $\hat{y}$ is obtained by a linear classification layer followed by a sigmoid activation:
$\hat{y} = \sigma(\mathbf{W}_r \ \mathbf{z}_{r} + b)$, with $\mathbf{W}_r \in \mathbb{R}^{1 \times 12d_c}$. The model is trained using the binary cross-entropy loss.

\section{Experimental Setup}

\textbf{\textit{Implementation Details.}}
TRENT and all competing methods are trained and evaluated on X-POSE under two experimental setups.
The primary setup uses a random split into 4,588 training, 559 validation, and 557 test samples. Performance is measured using macro F1 on the unfiltered test set and on high-agreement subsets ($h \ge 80\%, h \ge 90\%$). Applying thresholds $\theta \in \{10, 20, \dots, 90\}$ on this random test set results in pruned evaluation subsets of 494, 475, 448, 402, 362, 315, 269, 200, and 120 samples, respectively.
To evaluate temporal generalization, we introduce a chronological data split consisting of 4,563 training posts spanning 2017--11/2024 (with 95\% in 2023--2024), 570 validation posts from 11/2024--01/2025, and 571 test posts from 01/2025--04/2025.
Across both setups, the validation sets are used for hyperparameter tuning over the number of evidence items $M \in \{1,3,5,10\}$ and learning rate $\eta \in \{10^{-4}, 5\times10^{-5}\}$. 
All models are optimized with a batch size of 512 using Adam for up to 50 epochs, with early stopping after 10 epochs of validation stagnation. Random seeds are set to 0 across PyTorch, Python, and NumPy for reproducibility.

\paragraph{\textbf{Competing Methods.}}
As no prior work has evaluated on X-POSE, we implement a broad set of MFC models: \textbf{CCN} \cite{abdelnabi2022open}, \textbf{ERIC-FND} \cite{cao2025external}, and \textbf{ECENet} \cite{zhang2023ecenet} which leverage intra-modality evidence consistency, as well as \textbf{DT-Transformer} \cite{papadopoulos2023synthetic}, \textbf{RED-DOT} \cite{papadopoulos2025red}, \textbf{MUSE}, and \textbf{AITR} \cite{Papadopoulos_2025_WACV} that model all-to-all interactions within a single stream.
For a fair comparison, all replicated methods utilize a CLIP ViT-L/14 backbone and follow the training protocols and official repositories of the original papers.
Additionally, we evaluate several open-source and commercial VLMs for zero-shot detection on X-POSE, using \textit{Prompt~6} to incorporate the image-text pair alongside the extracted evidence excerpts: MiniCPM~\cite{yao2024minicpm}, Gemma~3 and 4, GPT-5~Mini and 5.4~Mini, Gemini~2.5~Flash and 3~Flash, Grok~4~Fast, and Claude~4.6~Sonnet. 
Except for MiniCPM (which is run locally), all VLMs are accessed via the OpenRouter.ai API.
Finally, we evaluate TRENT on NewsCLIPpings+ and VERITE, comparing it against optimized detectors \textbf{RED-DOT} and \textbf{AITR}, as well as VLM-based detectors \textbf{SNIFFER} \cite{qi2024sniffer}, \textbf{TRUST-VL} (LLaVA-1.5) \cite{yan2025trust}, \textbf{E2LVLM} (Qwen2-VL) \cite{wu2025e2lvlm}, \textbf{DEFAME} (GPT-4o) \cite{braun2025defame}, and \textbf{MAD-Sherlock} (GPT-4o) \cite{lakara2024mad}, reporting performance from their original papers.

\section{Results}

\begin{table}[t]    
    \centering
    \caption{Performance comparison on X-POSE. We report Macro-F1 across the random data split (including full test set and high-agreement subsets), alongside a past-only evidence setting evaluated on the full 
    ($h \ge 0\%$) random and chronological data splits.}

    \begin{tabular}{l ccc c c}
    \toprule
    & \multicolumn{4}{c}{\textbf{Random Split}} & \textbf{Chrono. Split} \\
    \cmidrule(r){2-5} \cmidrule(l){6-6}
    & & & & \multicolumn{2}{c}{\textbf{Past-Only Evidence}} \\
    \cmidrule(l){5-6}
    \textbf{Methods} & \textbf{F1} & \textbf{F1} ($h \geq 80\%$) & \textbf{F1} ($h \geq 90\%$) & \textbf{F1} & \textbf{F1} \\

    \midrule
    MiniCPM & 56.85 & 55.59 & 56.55 & 53.67 & 56.48 \\
    Gemma 3 & 54.48 & 56.16 & 54.92 & 54.02 & 52.81 \\
    Gemma 4 & 60.86 & \underline{65.91} & \underline{68.33} & \underline{60.31} & \underline{58.67} \\  
    GPT-5 Mini & 56.36  & 57.12 & 58.74 & 56.38 & 55.21 \\
    GPT-5.4 Mini & 57.48 & 57.37 & 57.91 & 54.73 & 55.49 \\
    Gemini 2.5 Flash & 56.65 & 62.49 & 63.96 & 56.18 & 54.42 \\ 
    Gemini 3 Flash & 61.33 & 63.32 & 64.91 & 59.93 & 56.86 \\   
    Grok 4 Fast & 57.67 & 62.99 & 64.91 & 57.54 & 57.67 \\
    Claude Sonnet 4.6 & 57.15 & 61.28 & 65.43 & 54.17 & 56.78 \\
    
    \midrule
    
    CCN & 58.61 & 61.45 & 58.52  & 56.89 & 54.33 \\
    ERIC-FND & 55.56 & 57.33 & 61.03 & 57.27 & 55.22 \\
    ECENet & 57.92 & 61.20 & 61.16 & 56.91 & 56.67 \\        
    
    MUSE & 53.36 & 55.36 & 57.27 & 53.76 & 49.37 \\                
    DT-Transformer & 57.99 & 60.78 & 62.25 & 57.43 & 57.94 \\
    RED-DOT & \underline{61.56} & 62.00 & 63.17 & 59.69 & 56.23 \\
    AITR & 58.46 & 64.02 & 63.78 & 57.99 & 56.77 \\
    
    \midrule
    
    TRENT & \textbf{63.10} & \textbf{67.36} & \textbf{70.83} & \textbf{62.22} & \textbf{60.14} \\        
    \bottomrule
    \end{tabular}
    \label{tab:sota}
\end{table}

\subsection{Comparative Results}

Table~\ref{tab:sota} presents the performance of TRENT on X-POSE, against a wide range of MFC detectors. 
Our model consistently outperforms all baselines across the standard random split, achieving macro-F1 scores of 63.10\%, 67.36\%, and 70.83\% on the full test set and high-agreement subsets ($h \geq 80, 90$).
Under a past-only evidence setting---only allowing evidence published before the user post---TRENT's performance drops to 62.22\% on the random split and 60.14\% on the chronological split. 
This decrease is expected given the significantly more challenging and realistic constraints of these settings; however, TRENT still consistently outperforms all competing methods.

Crucially, this competitive efficiency stems from architectural design rather than scale.
TRENT comprises only 5.13M trained parameters, significantly fewer than specialized methods like RED-DOT (16.9M) or AITR (18.7M), and orders of magnitude smaller than the tens or hundreds of billions of parameters of VLMs. 
Additionally, excluding the shared upfront costs of evidence excerpt extraction and embedding generation (which are identical across all models and performed only once), 
TRENT requires a mere 6.6--32.6 MFLOPs (scaling from 2 to 20 evidence documents) and is trained in just 21--30 seconds on a single RTX 3060 GPU. 
Furthermore, TRENT performs inference on the test set (557 samples) in less than 1 second; in stark contrast, evaluating the same inference set took 11m for GPT-5.4 Mini, 26m for Claude Sonnet 4.6, 33m for Gemma 4, and up to 56m for Gemini 3 Flash.

Within specialized MFC methods, all-to-all attention-based models (AITR, RED-DOT, DT-Transformer) tend to outperform intra-modality consistency models (CCN, ERIC-FND, ECENet). 
The latter restrict examination across modalities, missing key interactions between the claim and image evidence, and the image with textual evidence. 
Yet, while all-to-all architectures attend to all inputs simultaneously, their single-stream design can still overlook fine-grained dependencies. 
In contrast, TRENT explicitly models text $\rightarrow$ all evidence, image $\rightarrow$ all evidence, and internal image-text interactions through three dedicated cross-attention streams, combined with relational fusion, thereby better isolating subtle supporting or contradicting signals.

Among the evaluated VLMs, Gemma 4 and Gemini 3 yield the highest performance behind TRENT. 
We observe improvements within these model families, with Gemma 4 significantly outperforming Gemma 3, and Gemini 3 surpassing Gemini 2.5. While these gains may stem from enhanced reasoning capabilities, they could also be partially attributed to data contamination, as these large models could have encountered some of the target posts, events, or associated source articles during pre-training. 
Nevertheless, this risk is mitigated on the chronological split, which features minimal overlap with the training data of these VLMs due to its post-January 2025 timeline.

Moreover, as shown in Table~\ref{tab:comparison}, TRENT outperforms RED-DOT, E2LVLM, TRUST-VL, SNIFFER, DEFAME, and MAD-Sherlock across NewsCLIPpings+ and VERITE, while achieving competitive performance with AITR. 
However, specialized MFC models like RED-DOT and AITR struggle on X-POSE because they were developed on out-of-context benchmarks prone to retrieval shortcuts (Section~\ref{sec:mfc_datasets}), where the similarity-based baseline MUSE attains high scores on NewsCLIPpings (90.0\%) and VERITE (80.5\%) due to leakage \cite{Papadopoulos_2025_WACV}.
Similarly, when MUSE shortcuts are ablated, AITR's performance significantly decreases on VERITE, from 81.0 to 71.0\%. 
In contrast, MUSE drops to near-random on X-POSE (53.36\% on the full test set and 49.37\% on the chronological split). 
This sharp degradation indicates that X-POSE avoids such retrieval shortcuts.

\begin{table}[t]
\centering
\caption{Comparison on NewsCLIPpings+ and VERITE. 
The top two rows (shown in gray) denote methods relying on retrieval shortcuts.}
\label{tab:comparison}
\begin{tabular}{lcc}
\toprule
\textbf{Method} & \textbf{NewsCLIPpings+} & \textbf{VERITE} \\
\midrule
\textcolor{gray}{MUSE (MLP)} & \textcolor{gray}{90.0} & \textcolor{gray}{80.5} \\
\textcolor{gray}{AITR (w/ MUSE)} & \textcolor{gray}{93.3} & \textcolor{gray}{81.0} \\
\hline
RED-DOT & 90.3 & 76.9 \\ 
AITR (w/o MUSE) & 89.4 & 71.0 \\ 
SNIFFER & 88.4 & 74.0 \\
E2LVLM (Qwen2-VL) & 89.9 & 74.4 \\
TRUST-VL (LLaVA-1.5) & 90.4 & 73.6 \\
DEFAME (GPT-4o) & -- & 78.4 \\
MAD-Sherlock (GPT-4o) & \underline{90.8} & \underline{79.5} \\
\textbf{TRENT} & \textbf{92.5} & \textbf{79.6} \\
\bottomrule
\end{tabular}
\end{table}

\begin{table}[t]
    \centering
    \caption{Ablation analysis of TRENT's components.}    
    \begin{tabular}{rlccc}  
        \toprule
        \footnotesize
         \textbf{\#} & \textbf{Ablation} & \textbf{F1} & \textbf{F1 ($h\ge80\%$)} & \textbf{F1 ($h\ge90\%$)} \\
        
        \midrule

        1 & Trained on filtered data & - & 64.91 & 66.43 \\ 

        \midrule
        2& Full article features & 58.88 & 60.33 & 62.48 \\
        3& Relevant summaries & 60.33 &	65.99 &	67.20 \\
        4& Multi-E5 & 60.17 & 62.93 & 68.01 \\   

        5 & $-\text{Evidence quality filter}$ & 61.35 & 66.05 & 68.19 \\

        6 & $-\text{Evidence reranking}$ & 61.37	& 65.95 & 69.99 \\
        
        7 & $-\mathcal{I}$ & 57.62 & 58.42 & 58.32 \\ 

        8 & $- \mathcal{E}$ & 59.23 & 61.28 & 63.29 \\ 
        9 & $- \mathcal{E}^I$ & 61.75 & 61.48 & 64.04 \\ 
        10 & $- \mathcal{E}^T$ & 59.52 & 66.13 & 68.32 \\

        \midrule
        
        11 & Intra-modality Consistency $-$ $\mathcal{R}$ &  59.46 & 61.34 & 61.90 \\
        
        12 & Intra-modality Consistency & 58.83 & 62.25 & 62.13 \\

        13 & $- \mathcal{R} $ & 60.41 & 65.00 & 63.17 \\ 
        
        14 & $-\mathbf{c_{t \rightarrow e}}$ & 58.94 & 61.49 & 63.08 \\

        15 & $-\mathbf{c_{i \rightarrow e}}$ & 
        61.88 & 62.50 & 64.96 \\ 

        16 & $-\mathbf{c}_{t \leftrightarrow i}$ 
        & 58.50 & 62.34 & 65.83 \\        
        
        \midrule
        & TRENT & \textbf{63.10} & \textbf{67.36} & \textbf{70.83} \\ 

        \bottomrule
    \end{tabular}

    \label{tab:ablation}
\end{table}

\subsection{Ablation Study}

Table~\ref{tab:ablation} presents ablations analyzing the training data (Ablation~1), inputs (Ablations~2--10), and architectural components (Ablations~11--16) of TRENT.

\paragraph{\textbf{Scale vs.\ Data Quality.}}
All models are trained on the full training set and evaluated on the test set and two high-agreement subsets ($h \ge 80$, $h \ge 90$).
Training only on high-agreement data (Ablation~1) improves label reliability but reduces the dataset to 1,765 ($h \ge 80$) and 1,015 ($h \ge 90$) samples, causing a clear performance drop. For TRENT, scale outweighs consensus during training.

\paragraph{\textbf{Modality and Evidence Representation.}}
Ablations~2-4 modify evidence representations.
Using full-article embeddings from a long-context LLM (Ablation~2; 4096-dim Qwen GTE-7B\footnote{\url{https://huggingface.co/Alibaba-NLP/gte-Qwen1.5-7B-instruct}} with \textit{Prompt 5}) substantially degrades performance, suggesting verification cues are localized and diluted in full-document embeddings.
Replacing excerpts with VLM summaries (Ablation~3; MiniCPM with \textit{Prompt 5}) also reduces performance, likely due to loss of fine-grained details.
Substituting the CLIP-aligned text encoder with multi-E5 (Ablation~4) yields no gains, likely due to visual-textual misalignment.

\paragraph{\textbf{Evidence Filtering and Reranking.}}
Allowing low-credibility and social media sources in the evidence pool reduces performance (Ablation~5), validating the role of credibility-based filtering in mitigating the impact of noisy or biased evidence.
Similarly, disabling reranking (Ablation~6) results in a slight performance decline on high-agreement subsets. 
This indicates that while the evidence retrieval and excerpt extraction pipeline yields reasonably relevant candidates, reranking provides a beneficial refinement step.

\paragraph{\textbf{Contribution of Modalities and External Evidence.}}
Removing images (Ablation~7) drastically degrades performance, confirming the contribution of the visual signals and that X-POSE is not dominated by unimodal shortcuts. 
Removing all external evidence ($-\mathcal{E}$, Ablation~8) causes a significant drop, underscoring the necessity of external information.
Removing image-retrieved evidence ($-\mathcal{E}^I$, Ablation~9) is more detrimental than removing text-retrieved evidence ($-\mathcal{E}^T$, Ablation~10), suggesting that reverse-image search often provides crucial historical context, or provenance, missing from the user post.

\paragraph{\textbf{Evidence Triangulation and Relational Fusion.}}
Ablations~11-12 restrict processing to image $\rightarrow$ image-retrieved evidence and text $\rightarrow$ text-retrieved evidence, mirroring prior intra-modality evidence consistency architectures \cite{abdelnabi2022open, zhang2023ecenet}, and significantly reducing performance. 
Replacing relational fusion $\mathcal{R}$ (Ablations 11 and 13) with feature concatenation similarly weakens performance, underscoring its importance. 
Moreover, removing any individual stream from the three-stream architecture ($\mathbf{c_{t \rightarrow e}}$, 
$\mathbf{c_{i \rightarrow e}}$, 
$\mathbf{c}_{t \leftrightarrow i}$) (Ablations~14-16) significantly reduces performance.
Together, these findings validate TRENT’s design.

\begin{figure}[t]
    \centering
    \begin{subfigure}{0.53\linewidth}
        \centering
        \includegraphics[width=\linewidth]{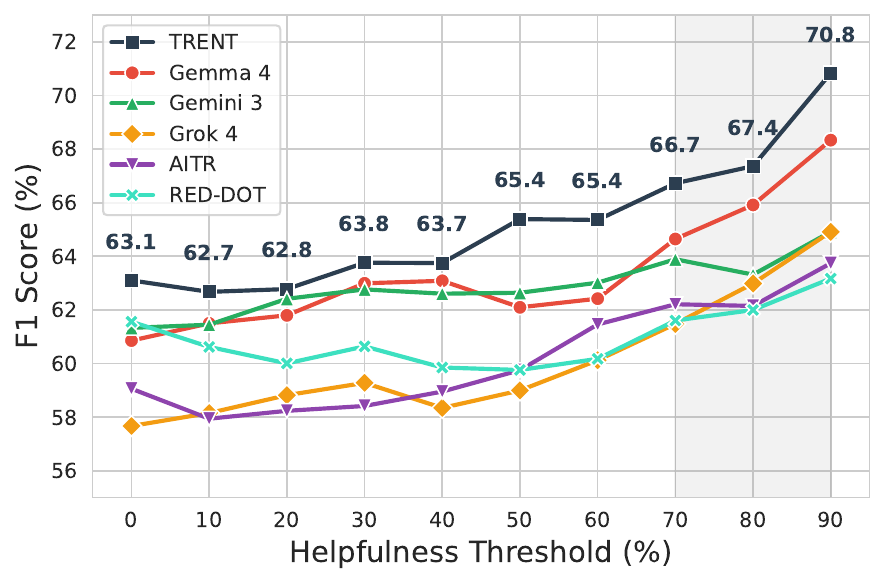}
        \caption{}
        \label{fig:helpfulness_theta}
    \end{subfigure}\hfill
    \begin{subfigure}{0.47\linewidth}
        \centering
        \includegraphics[width=\linewidth]{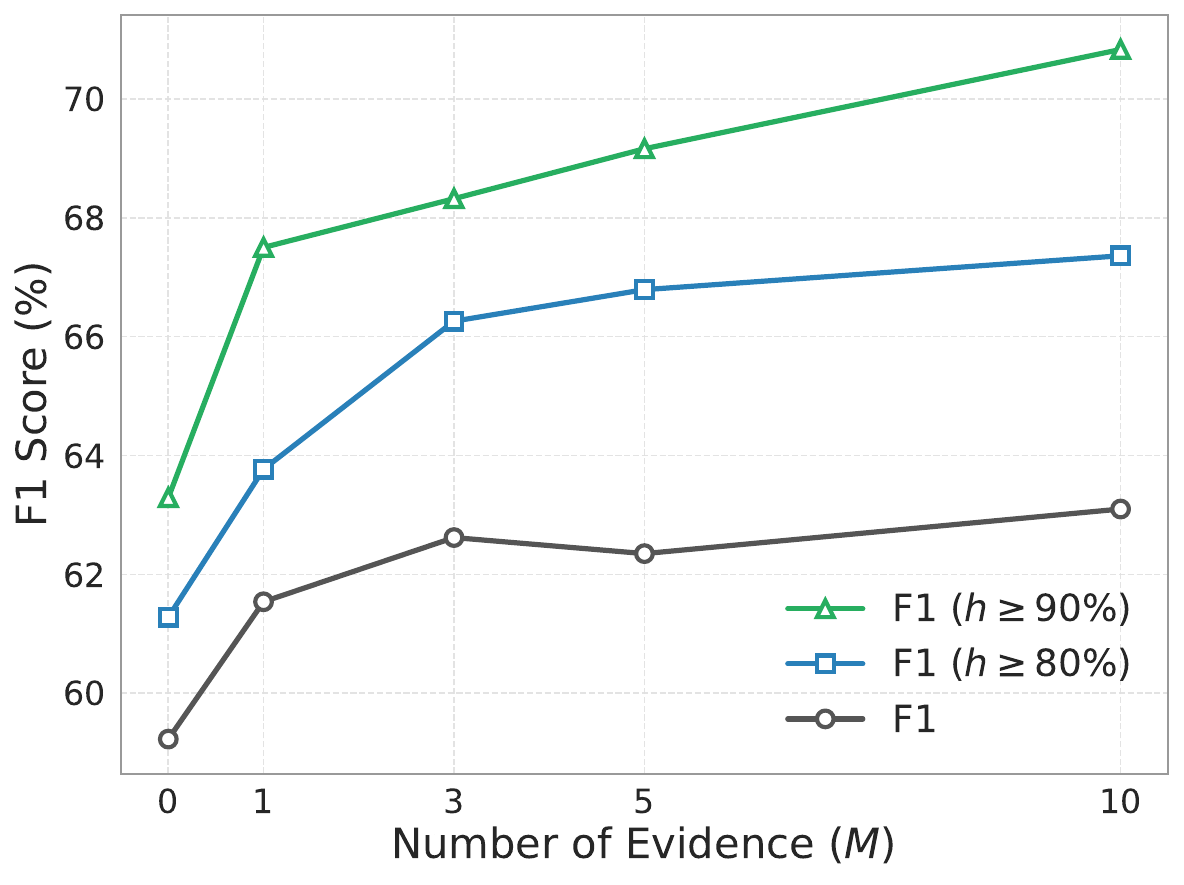}
        \caption{}
        \label{fig:topm_evidence}
    \end{subfigure}
    \caption{
    (a) Performance of TRENT and baselines across helpfulness thresholds $\theta$.
    (b) Impact of evidence quantity ($M$) on TRENT. 
    }
    \label{fig:tokm_evidence}
\end{figure}

\subsection{Further Analysis}

Figure~\ref{fig:helpfulness_theta} evaluates TRENT and five of the top performing methods across varying helpfulness thresholds $\theta \in \{10, 20, \dots, 90\}$. 
Performance remains static under low agreement ($\theta \le 60\%$, slope $\beta \approx 0.02$), suggesting that simple majorities (50--60\%) may suffer from user polarization. 
Conversely, beyond $\theta = 70\%$, methods exhibit a positive trend in macro-F1 ($\beta \approx 0.11$), indicating that super-majority agreement yields cleaner, higher-fidelity labels. 
However, enforcing these higher thresholds significantly reduces available sample volume, introducing a trade-off between annotation quality and dataset scale.

Figure~\ref{fig:topm_evidence} illustrates the impact of evidence quantity $M$ on TRENT. 
Performance improves markedly from $M=0$ to $M=1$ and continues to rise, peaking at $M=10$. 
This indicates that access to multiple evidence excerpts facilitates cross-source referencing, resolving ambiguities that a single excerpt ($M=1$) may miss. 
Furthermore, it demonstrates that TRENT can successfully isolate relevant signals from increasingly noisy, high-dimensional inputs.

\begin{figure}[t]
    \centering
    \begin{subfigure}[t]{0.49\linewidth}
        \centering
        \includegraphics[width=0.9\linewidth]{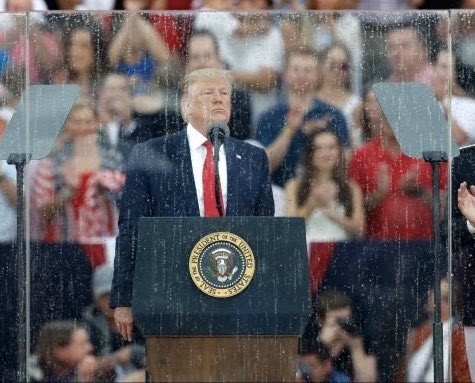}
        \caption{}
        \vspace{-5pt} 
        \begin{flushleft}\small
            \textbf{Post}: ``\textit{It looks like they froze him in an ice cube}'' \\
            \textbf{Community Note}: ``\textit{This image is dated July 5th, 2019...}'' \\
            \textbf{Label}: Misinformation \\
            \textbf{Prediction}: Factually Correct \\ 
            \textbf{Helpfulness}: 3.6\% (7 Helpful, 185 Not-helpful) \\
            \textbf{Excerpt}: ``\textit{The President delivered his speech behind a [...] bulletproof barrier.}'' [Source: CNN]
        \end{flushleft}
    \end{subfigure}\hfill     
    \begin{subfigure}[t]{0.49\linewidth}
        \centering
        \includegraphics[width=0.9\linewidth]{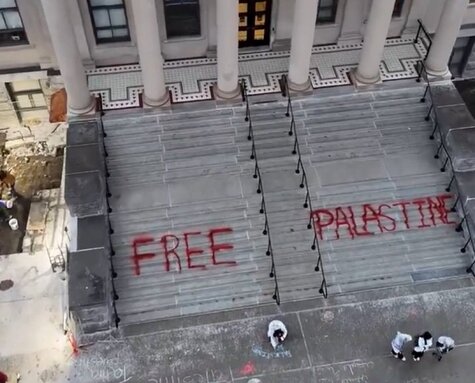}
        \caption{}
        \vspace{-5pt}
        \begin{flushleft}\small
            \textbf{Post}: ``\textit{Columbia Univ `smart' kids spell `Palestine' incorrectly...}'' \\
            \textbf{Community Note}: ``\textit{This is not Columbia... shows Tabaret Hall at U. Ottawa.}'' \\
            \textbf{Label}: Misinformation \\
            \textbf{Prediction}: Factually Correct \\
            \textbf{Helpfulness}: 100\% (6 Helpful, 0 Not-helpful) \\
            \textbf{Excerpt}: ``\textit{[...] at Columbia, [...], they can’t even spell Palestine.}'' [Source: Pacific Pundit]
        \end{flushleft}
    \end{subfigure}
    \vspace{-5pt}
    \caption{Qualitative examples of low consensus and evidence credibility.}
    \label{fig:errors}
\end{figure}

Figure~\ref{fig:errors} presents two inference examples that illustrate the importance of consensus-based and evidence-quality filtering. 
In (a), the retrieved evidence is credible, but the annotation is incorrect, as indicated by the very low helpfulness score (3.6\%), since the post is humorous rather than misinformation. 
This motivates our consensus-based filtering, which isolates higher-fidelity evaluation labels. 
In (b), despite a correct annotation (100\% helpfulness), the evidence originates from an unreliable, highly biased source\footnote{\url{https://mediabiasfactcheck.com/pacific-pundit/}} with low factual reporting, which provides misleading information that superficially supports the claim. 
This highlights the necessity of assessing evidence credibility (Section~\ref{sec:dataset}).

\section{Conclusion}

In this study, we introduced X-POSE, a novel \emph{in-the-wild} evidence-enhanced dataset for training and evaluating MFC models, comprising user posts annotated through X's Community Notes and augmented with VLM-refined external evidence. 
To address the limitations of prior MFC architectures, we proposed TRENT, which performs explicit evidence triangulation with relational fusion across text, image, and retrieved sources; it outperforms specialized MFC models and commercial VLMs, highlighting its efficiency and effectiveness.

Despite these improvements, overall performance remains limited, underscoring the difficulty of in-the-wild detection and the realism of X-POSE. This leaves ample room for further research. 
First, X-POSE can be used to develop and evaluate novel methods for in-the-wild MFC, while also exploring explainability \cite{warren2025show} by comparing model explanations against community notes. 
Second, because nearly 50\% of the retrieved articles in X-POSE currently lack credibility ratings, further research is required on evidence quality~\cite{glockner2022missing} and source credibility~\cite{srba2025survey}.
Finally, while our consensus-based filtering improves crowdsourced annotation quality by removing low-agreement entries, it introduces a quality-scale trade-off. 
To address this, X-POSE can be scaled up as new community notes are released, while expanding beyond X to other platforms like TikTok, Facebook, and Instagram to strengthen future research on cross-platform MFC.

\section*{Acknowledgments}
This work is partially funded by Horizon Europe projects AI-CODE and ELLIOT under grant agreement no. 101135437 and 101214398, respectively. 

\bibliographystyle{unsrt}  
\bibliography{main}

\end{document}